# Can machines perform a qualitative data analysis? Reading the debate with Alan Turing

Stefano De Paoli [s.depaoli@abertay.ac.uk]

## Abstract

This paper reflects on the literature that rejects the use of Large Language Models (LLMs) in qualitative data analysis. It illustrates through empirical evidence as well as critical reflections why the current critical debate is focusing on the wrong problems. The paper proposes that the focus of researching the use of the LLMs for qualitative analysis is not the method per se, but rather the empirical investigation of an artificial system performing an analysis. The paper builds on the seminal work of Alan Turing and reads the current debate using key ideas from Turing's "Computing Machinery and Intelligence".

This paper therefore reframes the debate on qualitative analysis with LLMs and states that rather than asking whether machines can perform qualitative analysis in principle, we should ask whether with LLMs we can produce analyses that are sufficiently comparable to human analysts. In the final part the contrary views to performing qualitative analysis with LLMs are analysed using the same writing and rhetorical style that Turing used in his seminal work, to discuss the contrary views to the main question.

## Introduction

The advent of Large Language Models (LLMs) or generative AIs (genAI) as some prefer to say, has brought increased attention about the possibilities to perform qualitative data analysis with these computing artifacts. This is a field that in just a couple of years went from a small number of rather primitive publications to several hundreds, few of which quite mature in terms of approaches and results. It is a field that has probably not yet found its 'centre of gravitas', and there are different positions, approaches and debate. One aspect that has been emerging lately is the pushback and rejection of the use of LLMs for qualitative data analysis by some qualitative scholars. This rejection debate rather than playing out on the ground of a healthy discussion based on empirical evidence, is moving forward often through normative and speculative arguments.

The manifesto of this is an open letter authored by Jowsey, Braun, Clarke, Lupton and Fine (Jowsey et al. 2025a) and signed by over 400 scholars. The authors argue for the rejection of LLMs (or genAI) in qualitative research. Albeit this is sometimes framed as a rejection for reflexive qualitative analysis (i.e. LLMs cannot exercise reflexivity), one of the key objections in the letter states very explicitly that "Q*ualitative research should remain a distinctly human*

*practice*" on the premise that "*GenAI as simulated intelligence is incapable of meaning making*". In other words, the letter makes broader normative claims about intelligence and knowledge that apply as a blanket to qualitative research. The letter, as it has been noted (De Paoli, 2025; Friese, 2025b), is not based on method related arguments, but on philosophical and epistemological assumptions. For example, it is based on the rejection of qualitative analysis which is not centred on a cartesian paradigm of the human cognition supremacy, and on the rejections of forms of cognition that are not purely human, and their capacities (existing or absent) to perform activities that are humans. As important as the philosophical problems are, they also are of limited value to assess whether it is practically possible to perform qualitative analysis alongside LLMs.

Critics also sought to evidence empirically the practical pitfalls of using LLMs. One example is the 'audit' work by Nguyen and Welch (2025) that has made a fair attempt to highlight some critical epistemic issues (I will return later to these). A key contribution to this literature is however a paper by Jowsey et al. (2025b) emphatically titled "*Frankenstein, thematic analysis and generative artificial intelligence*". What is interesting here is their empirical work. The authors sought to perform a thematic analysis using the Microsoft Copilot chatbot (as a LLM/genAI), on five Open-Access datasets, and to compare their findings both in terms of process and results with those of the original researchers that collected the data. Their results show something unequivocal: Copilot is simply not capable of performing anything looking like thematic analysis, let alone reflexivity. The authors therefore argue for caution, linking back to Shelley's trope of unleashing 'unknown powers'. In part this paper is more balanced than the manifesto letter rejecting LLMs (which has the same lead author), but is difficult not to see the thread: LLMs cannot perform anything looking like thematic analysis, both philosophical assumptions (Jowsey et al., 2025a) and empirical tests (Jowsey et al., 2025b) clearly demonstrate as much.

However, it is important, in the case of this empirical test that we perform a separate testing to confirm or disconfirm these findings. This is one of the purposes of this paper. Whilst Jowsey et al. (2025b) concentrated on five datasets, it is sufficient to focus on just one (Dunn, 2020), to understand their process and confirm or disconfirm their findings. The chosen dataset is sufficiently compact, and the limits of Copilot presented in relation to this data are well accounted for in their paper. Later, I will argue that the conclusions of Jowsey et al. (2025b) rest on a poor design of the empirical research.

In one of the most influential publications in the history of computing, Alan Turing (1950) famously posed the provocative question, "Can machines think?". In it, Turing also proposed what became known as the Turing Test, the design of an experiment used to evaluate whether a machine's responses could be indistinguishable from those of a human. Famously Turing replaced the question on whether machines can think with questions around the role of the machine in the empirical testing. The test is of course suggesting that if the response is indistinguishable then we could argue that the machine presents thinking capacities like humans. Building on this widely known idea, this paper reframes Turing's original challenge for the context of qualitative research. Instead of asking whether "machines can perform a qualitative (thematic) analysis?" (which is the problem that Jowsey et al. 2025b were posing, and other authors also contest) in a general sense, we can reformulate the problem as follows: "Could an artificial system carry out a qualitative analysis such that the results are close enough to that of a human analyst?". This question will allow us to investigate the boundaries of what is human and what is machine, and help us in raising fundamental issues about interpretation, meaning, and the role of human judgement in qualitative analysis with LLMs.

## The current debate

Alongside the popularisation of tools like chatGPT work has emerged attempting to perform qualitative analysis with LLMs. Early works were largely driven by engineering paradigms (Friese, 2025a), with computer scientists seeing this as a promising ground to test LLMs capacities at language manipulation (e.g. Chew et al, 2023; Xiao et al., 2023). This is still a significant area of work (e.g. Bano et al., 2025; Pattyn, 2025). For example, the title of a recent paper (Leça et al., 2025) recites: "*Applications and implications of large language models in qualitative analysis: A new frontier for empirical software engineering*". This leaves no doubt that these kinds of works approach the problem not from a social science perspective but from a technical, top-down, perspective. Alongside the engineering processes, part of the early narrative (which is still widely present in literature) was appropriated by the idea of delivering 'efficiency'. Qualitative analysis is known to be time consuming and using LLMs promises to reduce workload, save time and manual labour (e.g. Zhang et al. 2025; Parfenova et al., 2024; Jalali and Akhavan, 2024), or so goes this argument. This literature is part of the same class of engineering work, i.e. top-down efficiency.

Following however we have seen an increasing number of social scientists seeking to engage with the problem. Some studies were almost contemporary to the early engineering work. With these the focus was, in part, brought back to more familiar terrains of discourses around methods (i.e. methodology), empirical social sciences and in part was oriented at innovation in method practice. Some studies, including of the author (e.g. Ashwin et al., 2023; De Paoli, 2024a; Mathis et al., 2024; Wen et al., 2025), sought to replicate as far as possible (we could say imitate) key aspects of certain analysis methods, like the 6-phases of thematic analysis as proposed by Braun & Clarke (2006). Others sought to improve previously defined processes, for example with prompting techniques for thematic analysis (e.g. Khalid and Witmer, 2025), step-by-step instructions (Naem et al., 2025) or extension of standards (like COREQ) (Fehiring et al., 2025). Others attempted leveraging LLMs to bring some innovation in methods practice (e.g. Morgan, 2025; Freise, 2025a; Jenner et al., 2025; Heyes, 2025), diverging in part from established procedures, but still with intent of positively furthering qualitative methodological discourse. In this kind of literature there has not always been the intent to replace human-analysis with LLMs analysis.

The debate also encompasses several critical aspects. Discussions around bias, errors and limitations are of course very important (e.g. Ashwin et al., 2025; Jalali & Akhavan, 2024). The same is true for ethical issues (e.g. Schoroeder et al., 2025; Hayes, 2025; Davison et al., 2024), which clearly need to be at the centre of any method procedure, whether we use LLMs or not. Social sciences studies using LLMs, often had these concerns clearly in mind from the beginning. For instance, the study by Mathis et al. (2024) used a LLM (Llama) on a local machine to process sensitive data in ways that are compliant with data protection regulations.

Recently, we have seen a vigorous pushback from part of the qualitative research community. A paper by Nguyen and Welch (2025) dissected a number of studies and advanced several interesting observations, for example, about the epistemic nature of the results we can obtain from LLMs, and the extent to which they are comparable to human-centric epistemic knowledge creation. One of their concerns is that scholars seem to be too enthusiast about LLMs: "*By formulating more effective and convincing prompts, enthusiasts argue that users can enhance the relevance and accuracy of the model's outputs, thus increasing its utility in qualitative analysis*" (p. 26). In their conclusion the authors comment that "*This enthusiasm presents a profound epistemic risk to qualitative data analysis[...] because of the unwarranted trust placed in their capabilities to act as neutral or objective research instruments*" (p. 28). I will argue later in the paper that this position is due to a misunderstanding of what is the kind of research that some of us are attempting with LLMs.

Other studies do insist instead on the 'opacity' or 'black-boxed' nature of LLMs, like the one by Teixeira et al. (2025). Since we do not know what is going on inside ('the mind of') LLMs, we cannot trust their outputs. This is another epistemic angle, recurring across other studies. We cannot know if LLMs are really doing an 'understanding' and therefore the knowledge we receive is inherently fallacious. Again, I will show later that this assumption is moving from an incorrect reading of some of the work a few of us are performing. Beside many of the things that surround us have this black-boxed nature, yet we still trust them (e.g. flying in an aeroplane). This is a form of trust in expert systems (Giddens, 1990), which however we seem reluctant to award to LLMs.

The paper by Jowsey et al. (2025b), already described earlier, is another example. The authors performed an empirical investigation using five datasets to illustrate that the chatbot Copilot is simply not capable of producing anything valid. The authors here take the correct route (empirical test), however their design of the experiment is modest at best, and a significant part of the following pages will be devoted to illustrating this.

The letter by Jowsey et al. (2025a), lastly, is probably the apex of the critique, and this also was anticipated earlier. Some aspects of this letter will be rediscussed in the final part of this manuscript. The key argument is that human thinking is unique, and 'machine intelligence' cannot simply perform 'understanding'. Consequently, a qualitative analysis performed with an LLMs is 'metaphysically' not valid.

Table 1 summarises the four exemplar rejections of LLMs for qualitative analysis, there can be of course others, but these are the most prominent.

| Example rejection | Types of rejections (examples) | Claim |
|---|---|---|
| Joswey et al. (2025b) | Empirical | "Copilot (LLMs) cannot do a thematic analysis, here is the proof" |

| Joswey et al. (2025a) | Metaphysical | "Only humans can truly think, therefore a machine cannot do the analysis" |
| Nguyen and Welch (2025) | Epistemic (first type: enthusiasm) | "Machine analysis is an illusion of the enthusiasts" |
| Teixeira et al. (2025) | Epistemic (second type: 'opacity') | "We do not know what is going on inside, therefore machine analysis cannot be trusted" |

Table 1. Summary of rejections (examples)

## The Sciences of the Artificial

I argue here that the previous rejections are epistemically narrow positions that simply misunderstand the object under study. The object of study is not the method, which is tied with the idea of understanding. A more appropriate framing is to read the work on performing qualitative analysis with LLMs as an empirical study of artificial systems. This idea connects to Turing's original work. The Turing Test was never a metaphysical claim about 'what thinking is', nor an investigation of what is going on inside the 'brain of the machine', but rather a practical method for studying empirically how a 'general-purpose machinery behaves in interaction with humans'. Turing's intuition was that the best way to evaluate such an interaction was to observe it in a properly defined empirical setting.

Other authors have developed some of these ideas further. For example, Simon's (1996) emphasised that artificial systems (unlike natural systems) whether mechanical, computational, or organisational can be potentially studied empirically, and evaluated, in terms of how they operate in real-world contexts. Simon proposed an epistemological approach stating that the behaviour of designed systems is an empirical matter: we must study how inputs, and environments jointly shape the systems' actions. This requires gathering evidence not only about outcomes but also about processes. In this sense, LLM-based qualitative analysis is not the automation of human qualitative intellect, or a replacement of human qualitative analysis, but the investigation of how certain artificial systems behave when they are prompted to perform what we could consider an interpretive task. They may perform poorly or well, but we need to investigate this empirically.

It is crucial when we approach this empirical study that we perform a constant empirical refinement. Studying the artificial system means inspecting how they respond to variations in prompts, task structures, data inputs, human interventions and so on. We conduct the empirical investigation by adjusting inputs, observing failures, errors, and modifying the configuration until the behaviour finds some stability, at least in ways that we can consider 'good enough'. This is the type of methodological practice that some LLM researchers and qualitative scholars are undertaking, often without it being formally recognised.

Finally, it is important to note that humans themselves are a significant part of this artificial system. This is a hybrid. The Turing Test, for example, is based on the idea that humans are an active part of the design of the experiment, alongside many other things. It is the interaction between humans (the judge; the other human) and the computing machinery that generates the evidence on which the final judgement is based. This hybrid nature of the test was very well recognised for example by Latour (2008). In the same way, when researchers attempt to use LLMs for qualitative analysis, the 'human and the model' form a hybrid-system whose behaviour must be studied empirically rather than treated as the output of an isolated machine. When a researcher interacts with an LLM, for example through prompting and so on, the human and the model form some kind of hybrid cognitive arrangement (Mendoza and Quilantang, 2025). The system we are studying is therefore not the LLM in isolation, but the human–LLM assemblage that performs the analytic work.

The relevant question to ask therefore is not whether LLMs can deliver reflexivity, but rather how does this 'hybrid artificial system' behaves with what limitations, and evidence. Under this umbrella the current experimental work and much of the prior criticism can be situated in a clearer epistemic framework, one that acknowledges the novelty of the system under study and the necessity of investigating it empirically rather than normatively. I am convinced that this is an argument that also critics can accept.

## The 'Frakenstein' paper

We can now read back and indeed consider the empirical paper by Jowsey et al (2025b) using the previous framing. Theirs is an empirical study of the artificial system. Their process is more or less as described in Figure 1. They feed a very complex prompt to the LLM asking for all the 6 steps/phases of thematic analysis at once (plus more, requesting more information such as a trail of decisions, see prompt in Appendix). Moreover, they feed all at once an entire dataset of interviews/focus groups, of sizable amount. Lastly, they also remain fairly detached from the process and engage minimally with Copilot to see if they can refine the prompt, or indeed the empirical design to obtain better results.

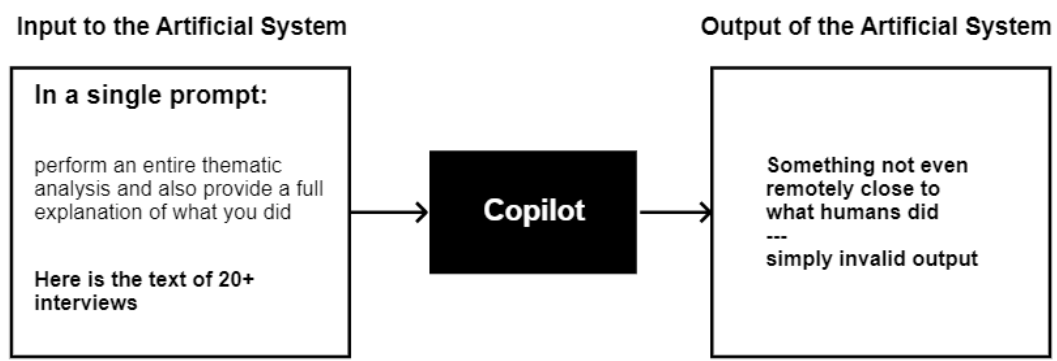

Figure 1. Experimental design of Jowsey et al. (2025b)

Even though the study is set-up as an empirical investigation, much of their focus then is not on building the process but on the output. The output they obtained is reported in a few tables (covering the five datasets) also using a comparative focus with the original published research. As anticipated, here I only focus on one of the studies, which is exemplar of their findings and will suffice for the purpose of illustrating the limitations of their design.

| Dunn et al. 2020 | Human researcher analysis | Copilot output (September 2024) |
|---|---|---|
| Number of participants identified | N=33 Parents (n=12) Practitioner (n=21) | Does not say |
| Themes | 10 plus 18 subthemes (covering both parents and practitioners, with overlap) | 8 themes, only about parents, some label overlap. No themes for practitioners. No themes across the 2 groups |
| Analytical process | Familiarisation by 2 researchers, 2 coders, framework, and final discussion with parents | No reflexivity. Opaque analytical process |
| Additional techniques | thematic framework and the index, iteratively created by 2 researchers, other reporting (e.g. COREQ) | Analytical process is opaque. Clear audit trail not provided |

Table 2. Findings from Jowsey et al (2025b) relating to Dunn et al, my summary

| Dunn et al. 2020 | Absolute score (n) | | Relative score (%) | |
|---|---|---|---|---|
| Quotes | Published results | Copilot results | Published results | Copilot results |
| Verbatim | 57 | 0 | 95.0% | 0.0% |
| Modified | 3 | 1 | 5.0% | 12.5% |
| Fabricated | 0 | 7 | 0.0% | 87.5% |

Table 3. Findings from Jowsey et al. (2025b) on quotes, Dunn et al.

The results obtained by Jowsey et al. (2025b) show that Copilot cannot produce the breadth of the themes produced by the human analysts, and additionally it only produces themes for only one of the two groups under investigation (i.e no themes for practitioners). There does not appear to be any reflexivity in the direct output of Copilot, because the LLM does not supply its 'underlying thinking' even though this is explicitly requested in the prompt, hence it is said to be 'opaque'. There is no self-audit of the process undertaken in the output of Copilot, by Copilot. Additionally, Copilot cannot tell how many participants were part of the research (Table 2).

A further part of the investigations relates to the quotes. This connects to wider debates about 'LLMs making things up', and their inability to 'understand' human language. When it comes to analysing the quotes used, the authors noted that Copilot tended to have almost only a few fabricated quotes. Moreover, if in the published paper we have 57 quotes reported from the underlying dataset, Copilot produced zero (Table 3).

These results are disheartening and would support immediately the idea that it is simply not possible to use Copilot (and by extension LLMs) to perform a meaningful analysis. However, these results are highly dependent on the way in which we study empirically the artificial system. I think that even just following three simple practical principles things can get much better than this.

First, even if a LLM is a 'black-box' there are things that we know about them after having tried to analyse data previously. For example, we know that they get overwhelmed by huge amounts of textual inputs. This is not a 'cognitive disability' of the LLMs. We have no idea if they will get better at that, it is simply a matter of how they are designed, based on current technical knowledge. Sending as input 20+ interviews in one single prompt is akin to try to fit 20+ people into a one-seater go-cart and expecting the cart will be still drivable. It is not a fault in the design of the go-cart, it simply is how the cart is, and poor judgment of the driver. It has been clearly addressed in previous work that there is a need to provide discrete amount of material for analysis.

Second, we also know that it is preferable to have the LLMs work on discrete tasks, performing steps separately. Again, this is not a 'cognitive disability', it simply is how they are right now. The prompt utilised by the authors asked Copilot all at once to do all the steps of thematic analysis. But this simply cannot work and will produce poor results. This is like wishing to prepare and eat a cake and putting all the raw ingredients together inside our mouth at once, instead of doing it in steps, first mixing eggs and flower, then cooking the cake, and eating it. This is like asking a scholar to perform all the steps of thematic analysis at once, instead of performing them in discrete phases.

Third, a proper empirical study would require a constant evaluation of the output and changes to the inputs, to work iteratively toward obtaining something which is at least good enough. Providing one (overcomplicated) input and judging the hybrid artificial system only one output is not sufficient. This would be like giving somebody a musical instrument and a complex set of instructions for how to play it and then judge their proficiency from the first sound they produce.

To summarise, these three aspects are elements which are essential, and mirror to an extent also what a human analyst would do: break down a full analysis (phases of thematic analysis), in each phase analyse a document at a time, go back to coding, themes and decisions to revise and improve the analysis.

**Redoing the analysis**

We can approach the empirical test using the Dunn (2020) dataset and the simple principles described above. For doing this I used the process I proposed in previous publications, simply because I am familiar with it and I have several scripts written in Python that I can easily reuse (this is my own reflexivity as part of the empirical design). The process is well documented and published, including in this journal. The source code used is Open Source and documented (De Paoli and Fawzi, 2025) (another reflexivity point). For the analysis I used the LLM GPT4o via Azure (December 2025) and the Application Program Interface (API). This is not the latest LLM, but it more than enough for the task at hand. I also requested and obtained permission from the UK Data Archive to use the model for data analysis.

In this approach, to perform initial coding, each interview is analysed independently, one at a time. For doing this I use a very simple zero-shot prompt (adapted from De Paoli, 2024). There is no evidence (as Jowsey et al, 2025b claim) that things could get better with more sophisticated prompts (like chain-of-thought), and in any case, we can see that even a simple zero-shot prompt, that was well-built over several iterations, can produce better results than asking an LLMs to perform an insurmountable task (see Prompt 2 in Appendix). I also normally use certain LLMs parameter such as temperature and top_p at zero in this stage, as I prefer some stability in the output, but this is my choice (reflexivity).

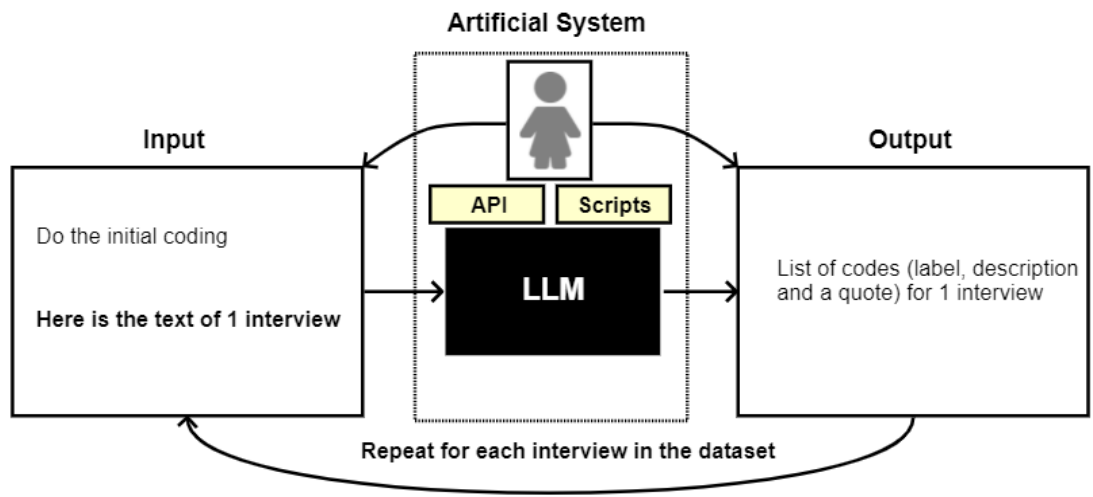

Figure 2. Artificial system and initial coding, example

Each interview has now an independent set of codes. However, we know there may be duplicates. Far from being an error this is something that happens also when humans do the coding, it means that the same code can be used to label portions of data across the same or different interviews. We have devised different approaches to deal with this, including few-shot prompts (e.g. De Paoli and Mathis, 2025b). We have also reflected on how this process has resemblance with some form of saturation of the initial codes (De Paoli and Mathis, 2025a).

Lately we also have proposed a different process for performing this saturation step, based on a pairwise comparison. This entails comparing pairs of interviews to identify unique and repeating codes. This is done comparing the label of the code as well as their descriptions. The resulting list of unique codes from each pair are further compared in pairs, until we arrive at just 2 pairs (list of codes) that are compared one last time to arrive at a final list of unique codes. The process keeps track of the decisions to merge two codes, of all of the associated quotes, and we can also ask the LLM to supply a brief description how why two codes were merged in this way (another possible component of reflexivity of the artificial system). The source code is Open-Source (De Paoli and Fawzi, 2025).

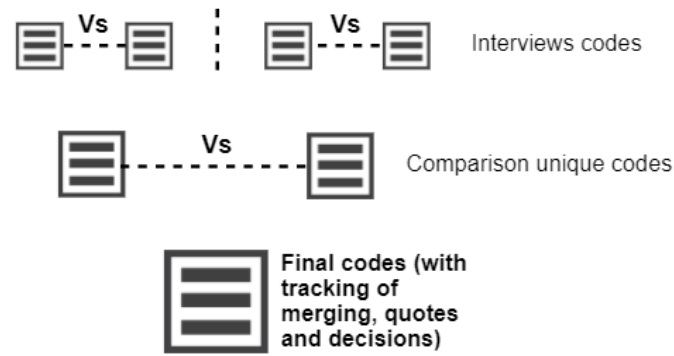

Figure 3. Finding saturation with pairwise comparison (simplified version)

Lastly with the list of initial codes it is possible to attempt at identifying themes. Again, good results can be obtained with simple prompts, designed over several experiments (e.g. prompt 3 in Appendix, adapted from De Paoli, 2024). We can ask the LLM the sort and group the codes into themes (either open-ended or a fixed number), using again the code name and their description. The analyst can also perform this multiple times to refine the themes. In this paper I also used temperature at 0.1, just to have the model work with some minimal creativity on themes.

## Results

We can now compare the results obtained with the above process to those obtained by Jowsey et al. (2025b). We can discuss each of their points in turn.

**The number of participants**

Jowsey et al. (2025b) requested Copilot to tell the number of participants based on their prompting. Why there is an expectation that Copilot can tell this information is unclear. As a researcher re-analysing an existing Open Access dataset, I can simply look at the files and count the participants. I can also read this information from the description of the data deposited on the UK Data Archive. This is really a non-argument, and I will not investigate it further.

**The Themes problem**

Joswey et al (2025b) contest that Copilot only produced 8 themes (against many more of the humans), and of these none do relate to the practitioners group. However, this result is simply a fallacy of their procedure. Using the process described earlier, I performed the analysis separately on: (1) the parents material and (2) the practitioners material and (3) on both set of data at once.

The analysis of the Parent material produced 146 codes (and related quotes), when duplicates are checked and merged the final number of codes is at 52 (still 146 quotes). There is therefore some saturation that we can easily calculate by the ratio of the two figures (i.e. 52/146). Table 4 present the themes generated, and in this case all codes were grouped within a theme.

| ID | name | description | Nr of Codes in the theme |
|---|---|---|---|
| Par1 | Emotional Challenges in Parenting | This theme encompasses the emotional struggles faced by parents with BPD traits, including feelings of inadequacy, anxiety, and the impact of mental health on parenting. It highlights how these emotional challenges affect their relationships with children and their overall parenting experience, providing insight into the complexities of their roles as caregivers. | 8 |
| Par2 | Support Systems and Community | This theme emphasizes the importance of community and support systems for parents with BPD traits. It explores how access to support, both from professionals and peers, can alleviate feelings of isolation and inadequacy, and enhance emotional well-being, ultimately impacting their parenting experience positively. | 8 |
| Par3 | Parenting Dynamics and Relationships | This theme focuses on the complexities of family dynamics, including the interactions between parents, children, and partners. It examines how differing temperaments, emotional responses, and external perceptions influence parenting styles and the overall family environment, shedding light on the relational aspects of parenting with BPD traits. | 8 |
| Par4 | Parental Advocacy and Education | This theme captures the efforts of parents to advocate for their children's needs, particularly in educational settings. It highlights the challenges they face in navigating systems designed to support children with special needs and the importance of intentional parenting strategies in fostering children's development. | 6 |
| Par5 | Coping Mechanisms and Strategies | This theme explores the various coping mechanisms employed by parents with BPD traits to manage their emotional pain and parenting challenges. It includes both positive strategies for emotional regulation and negative behaviors that may arise from their struggles, providing a comprehensive view of their coping landscape. | 7 |
| Par6 | Fear and Anxiety in Parenting | This theme addresses the fears and anxieties that parents with BPD traits experience regarding their children's well-being and their own parenting abilities. It highlights how these fears can lead to overcompensation and impact their parenting decisions, revealing the underlying emotional turmoil that shapes their experiences. | 5 |
| Par 7 | Honesty and Openness in Parenting | This theme emphasizes the importance of honesty and openness in parenting, particularly regarding mental health. It explores how parents navigate difficult conversations with their children about their struggles, aiming to foster understanding and emotional safety within the family. | 5 |
| Par8 | Impact of Mental Health Stigma | This theme highlights the stigma surrounding mental health issues and its effects on parents with BPD traits. It examines how societal perceptions influence their parenting experiences and the pressure they feel to conform to 'normal' standards, impacting their emotional well-being and interactions with their children. | 6 |

Table 4. Parents themes (artificial system)

Codes included in each theme are also available, for brevity I will show only one theme, just to reflect on the fact that the entire coding tree is indeed available from the analysis (Figure 3).

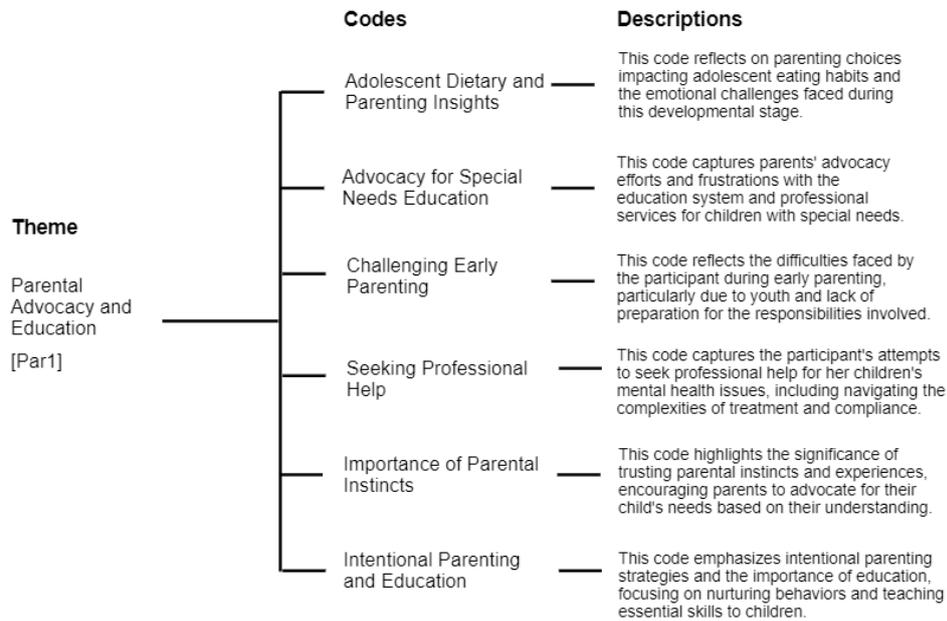

Figure 3. Example of coding tree we can obtain from the artificial system

For the practitioners the process produced 115 codes, and with the identification of duplicates 49 unique codes. Table 5 shows the themes, in this case 4 codes were not grouped in any theme, and this is normal as sometimes, even in manual analysis, some codes we produce do not necessarily fit with the patterns.

| ID | name | description | Nr of Codes in the theme |
|---|---|---|---|
| Prat1 | Emotional Challenges in Parenting | This theme captures the emotional difficulties faced by parents with BPD traits, including feelings of inadequacy, fear of judgment, and anxiety about child welfare. Understanding these emotional challenges is crucial for practitioners to provide effective support and interventions tailored to the unique needs of these parents, ultimately enhancing their parenting experience and child outcomes. | 8 |
| Prat2 | Support System Dynamics | This theme focuses on the complexities of support systems available to parents, including the importance of peer support, communication barriers, and the need for holistic assessments. It highlights how these dynamics influence parents' willingness to engage with services and the effectiveness of the support they receive, which is essential for understanding the broader context of parenting with BPD traits. | 8 |
| Prat3 | Parent-Child Relationship Focus | This theme emphasizes the significance of the parent-child relationship in the context of mental health support. It explores how parenting practices, emotional neglect, and parental mental health challenges impact child development and emotional well-being, underscoring the need for interventions that prioritize the child while addressing parental needs. | 7 |

| | Professional Support Challenges | This theme addresses the difficulties practitioners face when supporting parents with BPD traits, including balancing professionalism with care, navigating multi-agency collaboration, and dealing with entrenched core beliefs. Understanding these challenges is vital for improving practitioner training and enhancing the effectiveness of support services. | 7 |
|---|---|---|---|
| Prat4 | | | |
| Prat5 | Need for Tailored Interventions | This theme highlights the necessity for customized support strategies that address the unique emotional and practical challenges faced by parents with BPD traits. It underscores the importance of long-term support, mentorship, and preventative measures to foster resilience and effective parenting, ultimately improving outcomes for both parents and children. | 7 |
| Prat6 | Normalization of Parenting Struggles | This theme reflects the importance of normalizing the challenges associated with parenting, particularly for those with mental health issues. By acknowledging that difficulties are common, this theme promotes a supportive environment that encourages parents to seek help without fear of judgment, fostering a more inclusive approach to parenting support. | 8 |

Table 5. Practitioners themes (artificial system)

Contrary to what Jowsey et al (2025b) showed, we can therefore have themes relating to the practitioners, if we devise our empirical design in a proper way, following some basic practical principles. We can look back at some of the findings of both the original research and Jowsey et al. One of the complaints is that Copilot does not realise that Dunn et al. (2020) had themes overlapping between the two groups. Table 6 shows a different picture.

| Themes (Dunn) | Parents (Dunn) | Practitioners (Dunn) | Artificial hybrid themes (Parents) | Artificial hybrid themes (Practitioners) |
|---|---|---|---|---|
| Impact of mental health difficulties. 1.a. Emotional Intensity; 1.b. Coping strategy: Façade; 1.c. Coping strategy: control | X | X | Impact of Mental Health Stigma<br><br>**Emotional Challenges in Parenting**<br><br>Coping Mechanisms and Strategies | **Emotional Challenges in Parenting** |
| Impact of trauma. 2.a. Lack of parenting model; 2.b. Legacy of abuse. | X | X | --- | --- |
| Negative view of self as parent (partial). 3.a. Failure to live up to expectations; 3.b Stigma. | X | | Impact of Mental Health Stigma | |
| Unsupported parenting. 4.a. Social and family network; 4.b. | X | X | Support Systems and Community | Professional Support Challenges |
| Professional support. | | | | |
| Self in relation to child. | | X | | Parent-Child Relationship Focus |
| Lack of insight | | X | | Professional Support Challenges |
| Connection through shared understanding (partial). 1.a Alongside others with experience; 1.b Facilitation with understanding of mental health | X | X | --- | --- |
| Accessible not just available (partial). 2.a Logistics; 2.b Flexible in response | X | X | **Support Systems and Community** | **Support System Dynamics** |
| Support for children | X | | Parental Advocacy and Education | |
| Managing emotions | | X | | Need for Tailored Interventions |
| Normalizing parenting | | X | | Normalization of Parenting Struggles |

Table 6. Themes from the artificial system mapped against Dunn et al (2020)

Prima facie, the results do not look as bad as Jowsey et al, allude. Also, the mapping presented in Table 6 is just done based on theme labels and some of the descriptions, as I have understood them. More sophisticated mappings could be easily done (for example with semantic techniques, or with human experts, as we have done extensively elsewhere in e.g. Mathis et al., 2024), but as the authors simply inferred their results from the theme labels of their Copilot output, I should be allowed to play at the same level in this instance. It is the case then that in the analysis I performed anew on the same dataset, several themes may (more or less) be identified, mapping in some ways to those by Dunn et al., including across both groups of participants. Regardless we should not expect an exact match, because this is not like walking twice in the same river and expect to get in contact with exactly the same water as Jowsey et al. (2025b) themselves analogise in their paper. Differences are to be expected between different analysts.

A further complaint was that the LLMs cannot produce an articulation of themes and subthemes. However, this can be done. I performed the initial coding on both parents and practitioners material at once. This led to 273 codes (84 unique). We can use these to articulate themes and sub-themes. Again, we need to divide the process. We can use the LLMs to first generate a set of sub-themes (in this case I asked a fixed number close to the one by Dunn, where I counted 13, and Jowsey et al. 18, I asked for 14 in the prompt). Then in a second prompt we can work with the LLM to sort and group these into higher level themes (see Appendix prompt 4).

Moreover, one could manually decide from the first list which ones are already themes and decide to group the others as subthemes. Various scenarios are indeed possible. Table 7 shows a theme-subtheme articulation obtained, and the results are again not as Jowsey et al. (2025b) would suggest.

| Themes and subthemes (descriptions are omitted) |
|---|
| **Navigating Parenting Challenges** <br> Sub-themes: <br> • [0] Parenting Challenges and Skills <br> • [10] Barriers to Effective Parenting |
| **Understanding Mental Health Impact** <br> Sub-themes: <br> • [1] Mental Health and Parenting Dynamics <br> • [8] Child Development and Parenting Impact |
| **Building Support Systems** <br> Sub-themes: <br> • [2] Support Systems and Accessibility <br> • [7] Community and Peer Support |
| **Enhancing Emotional Communication** <br> Sub-themes: <br> • [3] Emotional Communication and Connection <br> • [11] Parent-Child Relationship Dynamics |
| **Addressing Trauma's Influence** <br> Sub-themes: <br> • [4] Impact of Trauma on Parenting <br> • [9] Coping Strategies and Resilience |
| **Navigating Parental Identity** <br> Sub-themes: <br> • [5] Parental Identity and Expectations <br> • [13] Reflections on Parenting Experience |
| **Fostering Professional Collaboration** <br> Sub-themes: <br> • [6] Professional Support and Collaboration <br> • [14] Navigating Professional Relationships |
| **Promoting Parental Education** <br> Sub-themes: <br> • [12] Education and Parenting Insights |
| **Strengthening Parent-Child Bonds** <br> Sub-themes: <br><br> • [15] Parenting Support and Connection |

| **Overcoming Barriers to Parenting** |
| :--- |
| Sub-themes: |
| • [10] Barriers to Effective Parenting |

Table 7. Articulation of themes and subthemes (artificial system)

I should also note that in this articulation there are sub-themes that complete the mapping I showed in Table 6. For example, there clearly is a theme about the 'impact of trauma'. Again, the results of this empirical test are not as described by Jowsey et al. (2025b)

**Framework and techniques**

The consideration that Copilot cannot provide a proper account of the process and there is no additional technique considered (Jowsey et al. 2025b) is also generally misguided. As I have detailed earlier there is reflexivity, the analysis followed key steps/phases, all detailed and accounted for, including Open-Source code. I supplied evidence for example that the coding tree is available. I have evidenced how the formulation of the codes/themes is inductive. In the next section I will also present quotes and all of them can be linked back to the data. All of these are available from the hybrid artificial system, and they can also be mapped against reporting standards like COREQ (where applicable), as show in Table 8. Again, this appears a non-argument based on the misunderstanding that Copilot alone (and with an overcomplicated set of instructions) can supply these techniques. Likewise, the expectation that Copilot can check the results with participant is misguided. First because what we are doing here is re-analysis of secondary data. Second because in a research based on primary data it is always possible for the human analyst to discuss the results with the participants.

| Topic | Item | Reflections[1] |
| --- | --- | --- |
| Number of data coders | 24 | 1 hybrid artificial system |
| Description of the coding tree | 25 | Provided (see example Figure 3) |
| Derivation of themes | 26 | Inductive |
| Software | 27 | GPT-4o, via API |
| Participant checking | 28 | Not applicable as this is secondary data analysis |
| Reporting Quotations presented | 29 | Yes (see next section) |
| Data and findings consistent | 30 | Yes. Several themes/sub-themes overlap with Dunn (based on preliminary observations as discussed, Table 6) |
| Clarity of major themes | 31 | Presented with full description (albeit no write up done, Tables 4 and 5) |
| Clarity of minor themes | 32 | Sub-themes presented and discussed against original author's themes |

Table 8 – Mapping of the analysis on the COREQ standard

---

[1] I am aware that COREQ would require here to report page numbers, but this is dependent upon the paper editing, and in any case a textual description here appears more suitable for the purpose.

**Quotes problem**

Through the initial coding process I described earlier (analysing together parents and practitioners), we obtained 273 initial codes (not-deduplicated) which means we have 273 quotes. With these I did an exercise to map which of these quotes are also in used in the paper by Dunn. For precision I excluded from Dunn's paper quotes with just 1 or 2 words. I used a very simple semantic similarity tool that I scripted in Python and then checked manually. Now there does not need to be a 100% match, after all quotes are identified during the analysis and they are a choice of the analyst.

There are quotes in the artificial system list of codes that are also used by Dunn et al. (2020) in their papers, I manually counted 20. Some examples are in Table 9

| Theme/Subtheme | Dunn's quote | Sub-theme/code | Artificial system extracted quote |
|---|---|---|---|
| Impact of Trauma [theme] | *"I can say most of my patients, most, maybe all of them have had horrific, awful lives. And then they just try and struggle through."* | Mental Health and Parenting Dynamics [sub-theme]<br><br>Code: Trauma Impact on Parenting | *"I can say most of my patients, most, maybe all of them have had horrific, awful lives. And then they just try and struggle through."* |
| Managing Emotions [theme] | *"I feel on one hand they would need a way to regulate their emotions, maybe like, uh, mindfulness course. Something that really, like, or skills course that really calms down the nervous system."* | Enhancing Emotional Communication [sub-theme]<br><br>Code: Parenting Emotional Communication Strategies | *"They would need a way to regulate their emotions, maybe like, uh, mindfulness course. Something that really, like, or skills course that really, ehm, calms down the nervous system."* |
| Lack of parenting model [subtheme] | *"Yeah, I think it's, you know, my mom was one end of the scale and I was at the other end, I think. There should have been some kind of middle."* | Reflections on Parenting Experience [sub-theme]<br><br>Code: Parenting Reflection and Experience | *"I think it was probably too much. Yeah, I think it's, you know, my mom was one end of the scale and I was at the other end, I think. There should have been some kind of middle."* |

Table 9. Quotes from the artificial system matching published quotes (examples)

On one occasion, upon manual scrutiny I identified a quote that was partially different between humans and the artificial system. Table 10 shows the original quote from the dataset, the quote from the published article and the one used by the LLM. What appears is that the LLM has used an ellipsis, it has tightened the quote, in a fashion that is similar to what is recommended when writing up results (see e.g. Lingard, 2019). The quote is not fabricated and it is still relevant. I should note that the original data had the ellipsis symbols (...) and it may very well be that the model did interpret that in the same way. This is the kind of behaviour we want to observe and, if we do not like, we can seek to improve via e.g. prompting. Moreover, the human analyst is always free to go to the original data and extrapolate the full quote. I should note that the quote in Dunn et al. paper truncated the original quote, added a comma and removed the ellipsis that was in the original data.

| Data | Dunn et al. | Artificial system |
|---|---|---|
| we want parents to attend group work at our group therapy program but there's never any childcare... or school holidays or whatever, I mean, that again *is a basic that if*- we need to be able to provide. | Then there comes the issue with we want parents to attend group work at our group therapy program but there's never any childcare, or school holidays, or whatever. I mean, that again is a basic. | We want parents to attend group work at our group therapy program but there's never any childcare... we need to be able to provide. |

Table 10. Use of ellipsis

To discuss further the findings from Jowsey et al. (2025b) I took a sample of the quotes from the 273 codes, precisely 21 codes[2]. To check if these quotes are in the data. All of them were traced verbatim in the original datasets. Of these 4 presented the case of ellipsis I discussed just above.

|  | In dataset | With ellipsis | In data | With ellipsis |
|---|---|---|---|---|
| Sampled quotes | 21 | 4 | 100% | 19% |

Table 11. Quotes from the artificial system also in the dataset

If we contrast these finds with those supplied by Jowsey (Table 3), we can see the stark difference, where they implied that all the quotes are fabricated. The analysis performed with a simple process and following some practical step produces quotes that are all in the data. Again, their argument rest on the LLMs incapacity to deliver valid outcomes but is clear that they results come from a modest design of the empirical test.

## Contrary Views on the Main Question

In his famous paper "Computing Machinery and Intelligence" Turing discussed several contrary views to the idea of whether machines can think. Most of those contrary views are formulated in a very similar way in literature that rejects the idea that with LLMs we can perform some kind of qualitative analysis. I would go over these contrary views and briefly present why they do not constitute a sufficient ground for a wholesale rejection. The reader will also allow me to use a 'Turing style' of writing in order to better align this part of the paper with Turing's original insights. I will also connect back to the observations I presented in Table 1 around the type of rejections that are formulated. The order in which I will present the views differs from that proposed by Turing, for practical and relevance reasons. I also will not discuss the contrary view related to Extra Sensory Perception (ESP), as this has not, as of yet, emerged in the debate and will replace this with a different objection.

*(1) The Argument from Consciousness.* This is a contrary view that often appears in the critiques about using LLMs in qualitative analysis. For example, Jowsey et al (2025a, p. 1) state: "*GenAI remains simulated intelligence only, based on statistical predictive algorithms without any understanding of the world, or the meaning of the language*". Originally for Turing,

---
[2] Statistically this would be 95% at +/- 20%, however we are not proving a hypothesis

this was simply an attempt of going after the wrong problem. It is not possible to simply ask if a machine can "really understand" or "really be conscious,", it is asking the wrong question. Consciousness and understanding are inaccessible black-boxes even in other humans. What we have here is confusing an empirical problem (how do we evaluate outputs of an artificial system, upon giving some clear and structured input?) with an ontological statement (the machine "lacks understanding of meaning"). One can see where the misunderstanding is from a quote cited earlier about "*the unwarranted trust placed in their* [LLMs] *capabilities to act as neutral or objective research instruments*" (Nguyen and Welch, 2025, p. 28). Critiques simply assume that LLMs are some sort of cartesian machines, that exist in their own solipsistic state of mind. They equate human brain to the LLMs (i.e. they assume anthropomorphism) and claim they are not the same thing. Whereas as I have shown earlier the process is much articulated and it requires a hybrid formation of humans and non-humans, to perform an analysis that we then study empirically. Another quote from the same paper is as follows: "*the essence of qualitative data analysis lies in the interpretation of meaning, an inherently human capability*" (p. 28). If an artificial system produces analyses, accounts, and justifications in ways that can meet accepted epistemic standards, then refusing to accept the analysis on the ground that something is inherently human is a category mistake. Nguyen and Welch move from lack of felt understanding to epistemic disqualification. Turing would insist we assess the problem through empirical criteria, and so should we.

*(2) Arguments from Various Disabilities.* For Turing these arguments take the form, 'I grant you that you can make machines do Y, but they will never be able do X'. He listed some examples like playing chess or compose arts. Recently a colleague of mine claimed that LLMs will never be able to understand "sarcasm" or "irony" when doing the analysis. I personally know many people who do not understand sarcasm or irony, but I never questioned if they are real people. Jokes aside, in Jowsey et al. (2025a) this is framed as follows "*We hold the position that only a human can undertake reflexive qualitative analytical work, and therefore use of GenAI is inappropriate in all phases… including initial coding.*" (p. 2), or elsewhere (Jowsey et al. 2025b, p. 9) "*GenAI systems do not possess autonomous cognitive functions… They lack the human capacity to interpret latent codes or uncover deeper meanings*" i.e. the machine will never be able to do X, they have disabilities. The problem with these statements as Turing noted is that they are based on some kind of impossibility claims, grounded not in empirical evidence but in some form of intuition. As I showed earlier a thematic analysis can be conducted which does not look that bad, and certainly it is not as bad as Jowsey et al. (2025b) want us to believe through their modest empirical test. But again, this is not a replacement for human analysis, it is an empirical investigation of an artificial system. Other forms of disability are imputed on an empirical ground "*Copilot outputs tended to draw most […] of its reported themes and quotes from the first 2–3 pages of textual data.*" (Jowsey et al, 2025b, p. 7), i.e. Copilot cannot do X. However, as we discussed earlier the problem is not that Copilot cannot do X, but rather that an empirical investigation has to be properly designed. This objection should be formulated as "researchers cannot do Y [e.g. proper design of the experiment], and as a consequence Copilot clearly cannot do X [because it received the wrong input]".

*(3) Argument of informality of behaviour.* In the original work by Turing this did relate to whether we could imitate informal behaviour with a computing machinery. There are traces of this in the contrary views relating to qualitative analysis, but they relate to the idea that qualitative work is "*deeply subjective*", "*complex*" and it deals with "*quirky*" and "*unpredictable*" voices and actions (Jowsey et al, 2025a). Put it differently an artificial system cannot capture the 'quirkiness of human action and intentions'. Framed in this way, this now looks like to fall under the idea of the disability 'the artificial system cannot do X' or, said it differently, 'we are informal, therefore machines cannot be like us'. However, this way of framing the problem shifts from showing empirically this impossibility to asserting this normatively, something we are needing to take for granted. Turing would say that empirical tests must judge whether artificial systems can approximate an understanding of the informal/quirky behaviour for the relevant epistemic tasks in ways that are at least 'good enough'. Until such tests are conducted and well documented we shall remain open that this could be possible.

*(4) Lady Lovelace's Objection.* This objection more or less states that machines just parrot back what we tell them to do, they do not create anything new. It is akin to Searle's Chinese Room experiment. Claims like an LLM is just "*simulating qualitative analysis*" (Jowsey et al., 2025b, p. 1) or "*only replicate patterns*" (Nguyen and Welch, 2025, p. 25) is a Lovelace-type of assumptions. However, as I illustrated earlier, we can perform an analysis with an LLMs and we ourselves are often quite surprised about the fact that the artificial system can identify themes and patterns that human researchers also have identified. In the end this is just a rhetorical argument and not an empirical one. If we were Turing, we would rather ask this problem: Does the system produce novel, informative, useful analytical insights? If yes, then claims about "mere simulation" are rhetorical rather than scientific.

*(5) The 'head in the sand objection'.* Turing frames this as follows: "*the consequences of machines thinking would be too dreadful*" that we should simply reject them based on some emotional ground. The paper by Jowsey et al. (2025b) proposing the Mary Shelly framing around 'Frankensetin' vaguely echoes this problem. In the critical literature this emotional tone is not explicit. However, I believe that where we can trace this issue is around the problem of 'opacity. Statements go as follows "*GenAI tools are inherently non-transparent … making their internal workings complex and the output they produce difficult, if not impossible, to explain*" (Nguyen and Welch, 2025, p. 6). This is some kind of 'epistemic discomfort'. When part of the artificial system's internal workings cannot be fully known, we may feel uneasy delegating interpretive capacities to the whole system. Hence, we are supposed to turn our head the other way and simply disengage. For example, Teixeira et al. (2025) note "*Such opacity around a central methodological tool is problematic in and by itself.*" (p. 3). It creates 'methodological discomfort', but again is this sufficient to suggest we should turn our head the other way? As I had noted in one of my earlier papers, it does seem inevitable that the qualitative research community will have to engage with LLMs, even if this was for seriously disqualifying their use, upon proper empirical evaluations. We should demand that this engagement is guided by clarity and proper empirical study, rather than epistemic discomfort alone.

*(6) The Theological objection.* In Turing's original paper this objection was relating to the idea of thinking and human immortal soul, a religious opposition so to speak. We cannot trace this, in this form, in the literature on qualitative analysis. However, we should acknowledge how sometimes critiques wish to frame the empirical research on artificial systems to perform qualitative analysis as "methodolatry", the worshipping of one method as the only truth. In an online post replying to comments about the manifesto letter, Lupton[3] had a rhetorical question "AI Worship: the new Religion". It may very well be there is people out there that treat LLMs as oracles, and those positions should be rejected. However, this argument of worshipping AI surely does not apply to the serious empirical investigation of artificial systems. The studies that out there engage in empirical research are driven by scientific curiosity and by the need for knowledge and understanding. The accusations of "methodolatry" simply point to a wrong problem. The object of study is not the method, but the empirical investigation of whether the artificial system can produce something valid.

*(7) The argument from hallucination.* I add one argument here to replace the EPS counter view of Turing. That of hallucination is a strange notion but as it has been noted one could think of it as the idea that "AI provides distorted information" (Ariso and Bannister, 2025). One can easily find this in the critiques, chiefly Jowsey at al. (2025b) noted how Copilot fabricated the few quotes it provided, surely that is distorted information. However, as I explained earlier that comes largely from the poor design of their empirical experiment. Beside the authors themselves made abundantly clear that also human analyses present troubling issues, as some authors fabricated quotes (not in Dunn et al., but a few of the other datasets they considered). On this ground we should abandon all qualitative analysis, but this is just a rhetorical point. For example, Jowsey et al (2025b) claimed that the paper by Mathis et al (2024) used chatGPT, whereas the authors used a model called Llama, deployed on a local machine. Is this another hallucination? The idea we should abandon the empirical study of the artificial systems capacity to perform an analysis based on the hallucination evidence supplied does not seem well grounded.

*(8) The Mathematical Objection.* This objection did relate to the idea that there are mathematical proofs that can "*be used to show that there are limitations to the powers of discrete-state machines.*" (Turing, 1950). This kind of argument is easily traced around ideas that LLMs are structurally limited: "*These limitations derive from the fundamental constraints of the transformer architecture and autoregressive method on which all current GenAI tools are built, and cannot be mitigated through technical improvements, prompt engineering, or scaling efforts.*" (Nguyen and Welch, 2025, p. 7). If Nguyen and Welch are claiming 'mathematical impossibility' (i.e., that no computational system could ever meet the interpretive standard), Turing would demand an empirical proof, that as of yet we have not seen, least of all the one supplied by Jowsey et al (2025b). There is a statement in Turing's paper that well capture this objection: "*Whenever one of these machines is asked the appropriate critical question, and gives a definite answer, we know that this answer must be*

---
[3] https://www.linkedin.com/posts/deborah-lupton-7ab43b270_people-who-say-that-those-of-us-who-reject-activity-7392679776303554560-Gek-/

*wrong, and this gives us a certain feeling of superiority. Is this feeling illusory?*". However, this feeling of human superiority is possibly misguided. First also humans make mistakes, as I have detailed in the previous objection, for example. Second, we tend to overestimate ourselves and underestimate machines because we have this tendency to selectively look at cases where machines fail (again the previous section on hallucination describes this, but also the section about the argument for disability, where a poor research design is used to blame the machine for failure). We are victims of some sort of Dunning-Kruger effect.

(*9*) *Argument from Continuity in the Nervous System*. In Turing's original paper he discussed the objection that human cognition is continuous, however a computing machinery is discrete, and this would imply that machines cannot really think like humans. A similar objection can be traced in almost all the critical papers, for example when we see claims like "*Discursive thematic analysis… is highly interpretive*." (Joswey et al., 2025a). Their critique rests on the assumption that human interpretive reasoning is continuous, holistic, and embodied, and thus inaccessible to discrete symbolic systems. We must be very sympathetic to this position, but this is again a philosophical assumption not an empirical one. It states that humans are a *res cogitans* and that LLMs are a *res extensa,* it is therefore an argument based on ontological dualism, and a typical cartesian move. The distinction between human "minds" and machine "artefacts" is a metaphysical prejudice, not empirical evidence, and from this, critics make descend both their empirical tests (Jowsey et al, 2025b) as well as their epistemic objections (Nguyen and Welch, 2025; Teixeira et al., 2025) which one could summarise as "interpretation requires a mind-substance, which machines lack.". Turing's response to this objection was that this difference does not preclude machines from imitating intelligence. We are not checking if "machines can think" in doing a qualitative analysis, but whether we can assess empirically if an artificial system can perform such an analysis.

## Conclusion

Can machines perform a qualitative analysis? I simply cannot provide a positive answer to the question. As I discussed earlier, this is not the right question to ask. However, what the previous pages showed is that we cannot simply accept a negative answer to the idea that an artificial system can produce a 'good enough' analysis. We rather should remain open to the possibility and investigate the problem empirically rather than basing a rejection on claims that are metaphysical, epistemic intuitions, or on misguided research designs. I myself do not think the artificial system would have done a better job than e.g. Dunn et al. (2020) in the analysis, but again this is looking at the wrong problem. The object of the study is not the method, but the empirical study of the artificial system. I would concur with the critics that there is work out there seeking to use LLMs which sees the method as the object of study, and they also are probably approaching the problem from the wrong perspective.

There clearly, and rightfully, is discussion about LLMs and qualitative analysis, and it is fully understandable that there is opposition too. However, it does appear that this has not been set up as a healthy debate. Philosophical assumptions can and should underpin methods (that is after all methodology), but reconciling opposing philosophical position seem hardly possible. If one group of scholars moves from a purely cartesian epistemology and another group moves from a relational ontology there is no reconciling, but it does not mean the second is necessarily wrong. If we must base rejection on the intuition that a machine is not like a human, then we are doing a poor service to science. Different is however showing each other the empirical evidence of research performed following some established methods. As I illustrated in the previous pages the current empirical evidence supplied as the basis for a rejection is not of enough good quality.

Ultimately the entire edifice of the rejection rests on considering LLMs as something objective, possessing a cartesian ontology in themselves which is evidently different from that of human beings. However, like for Turing the problem is not to demonstrate if machines can think (like humans) when doing an analysis, but to explore if an artificial system (human+machine) can approximate such an analysis.

## Appendix - Prompts

| Jowsey et al. 2025b prompt (1) |
| --- |
| Undertake thematic analysis of the dataset I have uploaded alongside this prompt. Review the dataset in its provided format. Begin by familiarizing yourself with the content and noting key recurring ideas. Generate initial codes based on identified features and apply these codes consistently across the dataset. Group related codes into potential themes, using pattern detection algorithms if applicable. Evaluate and refine the themes to ensure they are coherent and relevant. Develop detailed descriptions and clear names for each theme. Prepare a comprehensive report summarizing the thematic findings, supported by relevant data excerpts. Ensure consistency in coding and theme development, validate the accuracy of themes, and document the entire process transparently. Provide between four and eight themes, supporting the themes with the number of participants that are represented in each theme. Provide quotes from participants to support each theme. Provide this thematic analysis summary within 800 words. |

| Prompt for initial coding (2) |
| --- |
| Your task is to read the provided interview text and generate a comprehensive set of initial codes for thematic analysis, covering all the provided text. Focus your attention on capturing all significant explicit and latent meanings or events.<br><br>For each code, provide:<br><br>1. A meaningful descriptive name (maximum 5 words) |

2. A detailed description (50 words) explaining the code's meaning and relevance

3. A quote (minimum necessary to capture context and example, at least 150 words) from the text that exemplifies the code

Format the response as a JSON file with the following structure:

```
{
 "final_codes": [
   {
    "code_name": "Example Code Name",
    "description": "This is where you would provide a 25-word description of the code, explaining its meaning and significance in the context of the analysis.",
    "quote": "relevant quote here"
   },
   // Additional codes follow the same structure
 ]
}
```

Interview {one_interview_data}

| **Prompt for duplicates (3)** |
|---|

Compare the following target code against each comparison code and determine if they convey similar or the same meaning. Respond with a JSON object containing True/False values for each comparison.

Target Code:
```
{
  "code": "%s",
  "description": "%s"%s
}
```

Comparison Codes:
%s

Respond with a JSON object in this exact format:
```
{
  "comparisons": {
    "code_id_1": true_or_false,
    "code_id_2": true_or_false,
    ...
  }
}
```

LIST OF CODES: {List_of_all_codes}

| **Prompt for themes (4)** |
|---|

Read carefully the provided list of codes. Your task is to sort and group them into themes that capture the main ideas and patterns in the data, for a thematic analysis.

We are investigating the following research question: How do parents with Borderline Personality Disorder (BPD) traits experience parenting and how does this compare with the way their experience is conceptualized by practitioners who work with them?

For each theme:

1. Provide a meaningful theme name (maximum 5 words)
2. Write a detailed theme description (50-75 words) explaining the theme's meaning and relevance for answering the question
3. List the codes (by index number) that are associated with this theme

Important:

- It is okay for themes to have shared codes; this will facilitate an evaluation of higher-order concepts.
- Ensure that every code is assigned to at least one theme. Do not leave any codes unassigned.

Format the response as a JSON file with the following structure: [Json omitted]

LIST OF CODES: {list_of_unique_codes}

**Prompts for subthemes (5)**

Read the provided list of sub-themes and sort and group all of them into 10 themes that capture the main ideas and patterns in the data. Note that a sub-theme can only appear in one single theme, no duplicates.

We are investigating the following research question:

How do parents with Borderline Personality Disorder (BPD) traits experience parenting and how does this compare with the way their experience is conceptualized by practitioners who work with them?

For each theme:

1. Provide a concise theme name (maximum 5 words), use a verb/gerund to capture action

2. Write a detailed theme description (50-75 words) explaining the theme's meaning and relevance

3. List the sub-themes (by index number) that are associated with this theme

Format the response as a JSON file with the following structure [Json omitted]

SUBTHEMES: {List_of_subthemes}